\documentclass[11pt,twoside]{article}
\usepackage{asp2010}

\resetcounters

\begin{document}

\title{Interferometry and the Fundamental Properties of Stars}
\author{Guillermo Torres
\affil{Harvard-Smithsonian Center for Astrophysics, 60 Garden St., MA 02148}
}

\begin{abstract}
For many decades the determination of accurate fundamental parameters
for stars (masses, radii, temperatures, luminosities, etc.) has mostly
been the domain of eclipsing binary systems.  That has begun to change
as long-baseline interferometric techniques have improved
significantly, and powerful new instruments have come online.  This
paper will review the status of the field, and in particular how the
knowledge of precise stellar properties helps us understand stars.
Main-sequence stars similar to the Sun are by far the best studied,
but much remains to be done for other kinds of objects such as
early-type as well as late-type stars including brown dwarfs, evolved
stars, metal-poor stars, and pre-main sequence stars.  Progress is
illustrated with several examples of how interferometry has
contributed significantly in some of these areas.
\end{abstract}

\section{Introduction}

The determination of the fundamental properties of stars, i.e., their
masses, radii, effective temperatures, luminosities, etc., is a
classic discipline in Astronomy that goes back more than a century.
To many, this area of work may not seem as appealing or fashionable as
other topics such as cosmology and extrasolar planet research, which
seem to garner most of the attention these days. It is worth keeping
in mind, though, that there are few areas in modern Astrophysics that
do not rely to some extent on our knowledge of the basic properties of
stars, and this includes cosmology and extrasolar planets.

Perhaps the most important application of precise measurements of the
mass and other stellar characteristics is to improve our understanding
of stellar structure and stellar evolution, by comparing the
observations against predictions from current models. Numerous
examples of such comparisons may be found in the literature
\citep[see, e.g.,][]{Pols:97, Lastennet:02, Hillenbrand:04}, and have
shown that while our knowledge of the physics of stars appears to be
in reasonably good shape for solar-type stars, the agreement in other
parts of the H-R diagram is not as good. More practical applications
include the estimate of the total mass in stellar clusters making use
of the well-known mass-luminosity relation determined empirically from
observations of binary systems (the only ones allowing a dynamical
measurement of the mass). Binary stars with well-determined properties
serve also as valuable distance indicators, not only in the Milky Way
but also in other galaxies (SMC, LMC, M31, M33), and have become an
important tool for establishing the large-scale structure of the
Universe.

In recent years the field of extrasolar planets has highlighted the
importance of understanding stars. For transiting exoplanets, for
example, in which the observables are the transit light curve and
typically also a spectroscopic radial-velocity curve, the masses and
radii of these objects cannot be determined independently of those of
the parent star. Even if the inclination angle is precisely known in
these cases, the value of the planet mass derived from the
spectroscopic orbit scales as $M^{2/3}$, where $M$ is the mass of the
host star. Similarly, the measurement of the depth of the transit
events does not immediately yield the radius of the planet, but only
the \emph{ratio} between the planet radius and the stellar radius. It
is essential to know the properties of stars in order to learn about
planets. Thus, in this era of deep questions about the possibility of
life on other worlds, and the frequency of planets like our own in the
Universe, stellar Astronomy has become relevant again.

In this review I will outline the techniques applied to measure stars,
and the status of the field of fundamental stellar parameter
determination. I will also illustrate some of the contributions from
long-baseline interferometry, which is the subject of this Workshop.

\section{Methodologies}

The observational and analysis procedures for determining basic
stellar properties are well known, and I will only summarize them
briefly here, pointing out some of their advantages and limitations.

\paragraph{Masses}
A common misconception is that measurements of a binary system of any
type automatically yield dynamical masses for both components.  This
is of course not true. Only binaries with certain kinds of
measurements allow the individual masses to be determined without
assumptions.  Historically the most precise mass determinations have
come from double-lined spectroscopic binaries that undergo eclipses.
Note that \emph{single-lined} eclipsing binaries \emph{do not} yield
the individual masses; radial velocities need to be measured for both
components in order for this to be possible, unless one is willing to
\emph{assume} a value for the mass of one of the stars. And of course,
without eclipses neither single- nor double-lined spectroscopic
binaries allow the absolute masses to be calculated; only a lower
limit on the secondary mass can be derive in the first case, and lower
limits for both stars in the second case.

Another common way in which masses are derived when eclipses do not
occur is in double-lined spectroscopic binaries that are spatially
resolved, and in which the astrometric orbit of the secondary relative
to the primary is determined interferometrically or by other means
such as direct imaging \citep[for a review, see][]{Torres:04}. In this
case one can calculate not only the absolute masses of both stars, but
also the ``orbital parallax'' from the ratio between the
(de-projected) linear semimajor axis from spectroscopy and the angular
semimajor axis from astrometry. This added bonus makes double-lined
astrometric-spectroscopic binaries particularly useful because they
allow the individual luminosities to be computed from the apparent
brightness of the stars; thus, both the masses and the luminosities
can be known in these cases, independently of any models. The high
precision of long-baseline interferometric observations of some
binaries has yielded very precise distances and luminosities for
stars, in addition to the masses. If radial velocities are measured
for only one of the components, however, then an additional piece of
information is needed to obtain the individual masses. This usually
involves \emph{independent} knowledge of the trigonometric parallax,
such as from the Hipparcos mission \citep{Perryman:97,
vanLeeuwen:07}. If only the relative astrometric orbit is known, and
no velocity measurements are available, individual component masses
cannot be obtained without assumptions. In these cases one can only
infer the \emph{total} mass (from Kepler's third law), provided the
distance to the system is known.

Long-baseline interferometric measurements of binaries are usually
relative in nature, and require radial velocities (of both stars) to
be able to infer the masses. When the astrometric motion of each star
can be measured separately against a background of reference stars,
then spectroscopy is not needed at all to infer the masses of both
stars. An example of such absolute astrometric measurements and mass
determinations using the Fine Guidance Sensors on HST may be seen in
the work by \citet{Hershey:98}.

\paragraph{Radii}
Double-lined eclipsing binaries have been our main source of accurate
absolute radii for stars. The solution of the light curves yields the
individual radii in terms of the semimajor axis, and the spectroscopy
provides the absolute scale.  However, in recent years very precise
radius measurements have also been possible with long-baseline
interferometry (yielding angular diameters) combined with accurate
parallaxes from the Hipparcos mission. This is discussed further
below, in connection with low-mass stars. One difficulty that should
be pointed out is that angular diameters are usually only available
for single stars, and are typically very difficult to measure
interferometrically in binary systems. This is unfortunate, as
dynamical masses can only be measured in binaries. I will not discuss
here the subject of asteroseismology, which can also provide an
accurate measurement of a quantity closely related to the stellar
radius which is the mean density of the star. Reviews on the power of
asteroseismology for determining stellar properties can be found
elsewhere in this Volume.

\paragraph{Temperatures}

Interferometry enables the determination of effective temperatures for
stars in the most fundamental way, through the well-known relation
between the bolometric flux of a star as observed from the Earth
($f_{\earth,\rm bol}$), the temperature ($T_{\rm eff}$), and the
limb-darkened angular diameter ($\theta_{\rm LD}$). This relation is
$f_{\earth,\rm bol} = \sigma T^4_{\rm eff} \theta_{\rm LD}^2/4$, where
the symbol $\sigma$ represents the Stefan-Boltzmann constant.  Flux
measurements require observing the star at a range of wavelengths, and
fitting the spectral energy distribution with some model. The
dependence of the results on this model is minimal, however. More of
these determinations have become available in recent years thanks to
improvements in the precision of the interferometric angular
diameters.

Other ways of determining stellar temperatures are less fundamental.
For example, spectroscopic techniques rely heavily on model
atmospheres, and do not work very well for cool stars, which display
strong molecular features that the models still have difficulty
reproducing in detail. A common way of inferring temperatures
indirectly is through the measurement of a color index and the use of
color-temperature calibrations \citep[e.g.,][]{Casagrande:08,
Casagrande:10}. These are available in a variety of photometric
systems. Much progress has been made recently in understanding the
systematics that have affected these relations in the past, and
remaining biases are now believed to be smaller than 100\,K.
Photometric temperatures are sensitive to reddening, however, so care
is required with this approach.

\paragraph{Luminosities}

In eclipsing binaries in which color indices on a standard system are
available for both components, bolometric luminosities are easily
obtained from the absolute radii and photometrically estimated
temperatures through the Stefan-Boltzmann equation, with the same
caveat on the reddening as above.  For single stars, the widely used
Infrared Flux Method (IRFM) can provide both the angular diameters and
the temperatures at the same time \citep[see, e.g.,][]{Blackwell:77,
Blackwell:79, Ramirez:05}. When combined with a measurement of the
parallax, it is then possible to infer the luminosity. In practice
luminosities are often derived from a measurement of the apparent
brightness, a parallax (either trigonometric or ``orbital''; see
above), and bolometric corrections from standard tables. Because of
the arbitrary nature of the zero point of these corrections, which
many investigators often overlook, attention to this matter is
required to ensure consistency \citep[see][]{Torres:10a}.

\paragraph{Metallicities}

When the goal is to use measured stellar properties ($M$, $R$, $T_{\rm
eff}$) to test models of stellar evolution, knowledge of the chemical
composition of a star provides for a much more stringent comparison.
Otherwise this becomes a free parameter that can be changed in the
models so as to achieve the best fit. Metallicity is usually
determined spectroscopically through a spectral synthesis approach or
by measuring equivalent widths, although this is always more difficult
in binary systems with composite spectra. As for the case of
spectroscopically determined temperatures, poorly known molecular
opacities in the atmospheres of cool stars make metallicity
determinations problematic. Photometric calibrations for solar-type
and also cooler stars are available as well
\citep[e.g.,][]{Holmberg:07, Twarog:07, Schlaufman:10}, but
spectroscopic determinations are preferable when possible.

\begin{figure}[!t]
\plotfiddle{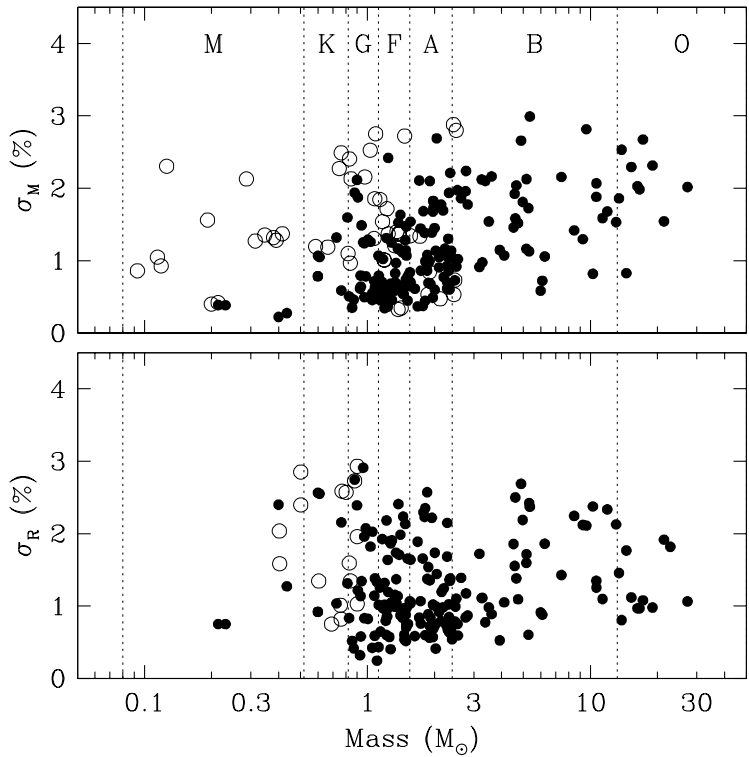}{9.0cm}{0}{60}{60}{-185}{-100}
\caption{Stellar mass and radius uncertainties as a function of mass
for the 94 eclipsing binary systems plus $\alpha$~Cen (filled circles)
with relative errors smaller than 3\%, from \cite{Torres:10b}. Open
circles in the top panel correspond to additional systems with mass
determinations from long-baseline interferometry (see text). Open
circles in the bottom panel correspond to stars with angular diameter
measurements from interferometry \citep[but with no corresponding
dynamical masses, which are estimated here using the mass-luminosity
relations of][]{Delfosse:00}. Spectral types are indicated along the
top.\label{fig:mr}}
\end{figure}

\section{Status of accurate mass and radius determinations for normal stars}

Binary stars have been studied for more than two centuries, yet the
number of eclipsing systems with accurate mass and radius
determinations represents only a tiny fraction of the many thousands
of systems known. An early review of the status of the field by
\cite{Popper:67} listed only two systems with absolute masses (but no
radii) known to 3\% or better, among many others with more poorly
determined properties.  An update 13 years later increased the tally
of well known systems to seven, this time including the radii
\citep{Popper:80}. Starting in the 1970's, efforts by the Danish group
led by J.\ Andersen brought the number up considerably, and by the
time of the next major review on the subject a total of 45 binaries
(90 stars) had masses and radii with relative errors of 3\% or better
\citep{Andersen:91}. Since then significant improvements have been
made in both the observational and the analysis techniques. As a
result, the number of well studied systems has more than doubled to 95
in the most recent review \citep[94 eclipsing systems, and
$\alpha$~Centauri;][]{Torres:10b}, which for the first time includes
an extragalactic binary.

Figure~\ref{fig:mr} shows the distribution of these systems as a
function of stellar mass (filled circles), with the relative errors in
mass and radius plotted on the vertical axes. In both panels the
slightly rising lower envelope toward higher masses is mostly due to
increasing difficulties with the spectroscopic analysis in early-type
systems, caused by strong winds and other complications in their
spectra.

Long-baseline interferometry has been making steady progress over the
last decade or so, and now contributes significantly to the list of
high-quality mass measurements. In fact, in addition to the eclipsing
binaries, the review by \cite{Torres:10b} lists some two dozen
interferometric binaries in which both components have mass
determinations that are also good to 3\% or better. These are shown
with open circles in the top panel of Figure~\ref{fig:mr}. While there
are very few well-studied eclipsing binaries among the later type
stars, interferometry is seen to be quite complementary and has added
a significant number of K- and M-type stars. The downside, as
mentioned earlier, is that these interferometric mass measurements are
not accompanied by the corresponding radius measurements, so are
generally of more limited value for testing models of stellar
evolution.

\section{Contributions of long-baseline interferometry}

In this section I describe a few selected areas in which
interferometry has made especially interesting contributions to the
determinations of the global properties of stars, and/or has provided
useful tests of stellar theory.  Other articles in these Proceedings
describe additional stellar quantities that interferometry is
particularly well suited to measure.

\subsection{Stellar radii and the discrepancies with models for low-mass stars}

As advanced as our current knowledge is of stellar structure and
stellar evolution, there are plenty of indications that our
understanding of stars is far from complete. A prominent example is in
the area of low-mass stars. For nearly four decades there has been
mounting evidence that the measured radii of these stars are larger
than theory predicts by up to $\sim$10\%, and also that their
effective temperatures are cooler than anticipated by up to
$\sim$5\%. Early indications of these anomalies were reported by
\cite{Hoxie:73} and \cite{Lacy:77}, and were strengthened by
\cite{Popper:97}, \cite{Clausen:99}, and others. More recent highly
accurate determinations of the masses, radii, and temperatures in
low-mass eclipsing binaries such as YY~Gem, CU~Cnc, GU~Boo, CM~Dra,
and others have now removed all observational ambiguity
\citep{Torres:02a, Ribas:03, Lopez-Morales:05, Morales:09}. Enlarged
radii have been confirmed in a number of additional systems, although
I note that these measurements are not all equally reliable. Many of
the mass-radius diagrams seen in the recent literature show
considerable dispersion, making the picture rather confusing. Some of
this scatter may be due to real differences between systems, but it is
likely that the published uncertainties for many of these binaries are
unrealistically small, and do not account for systematics. In
particular, few of these studies document any tests to investigate the
effects of spots, which are prevalent in late-type stars \citep[see,
e.g.,][]{Morales:08, Windmiller:10, Morales:10}, or offer external
constraints as a check on the light curve solutions.

In any case, it is clear that current stellar evolution models
underestimate the radii of late-type stars, and overestimate their
effective temperatures.  The effect is believed to be due to
chromospheric activity in close binary systems. Indeed, most of the
stars that show this anomaly are in short-period binaries, where
strong tidal forces drive the components into spin-orbit
synchronization (rapid rotation). As a result, these systems often
display variability due to spots, X-ray emission, and spectral
signatures of activity such as \ion{Ca}{ii} H and K emission,
H$\alpha$ emission, etc.\ \citep[see, e.g.,][]{Lopez-Morales:07}.
While there is in fact a theoretical understanding of the impact of
activity on the radii and temperatures of low-mass stars \citep[][and
others]{DAntona:00, Mullan:01, Chabrier:07}, these effects are yet to
be incorporated into publicly available stellar evolution models.

An obvious action item on the part of observers is to now focus on
long-period low-mass eclipsing binaries, in which the stars might be
expected to rotate more slowly and to therefore be less
chromospherically active. These could show better agreement with
theory, thereby supporting the idea that activity is the culprit.
Although a few such systems have been found from the ground, they are
generally rare and challenging to study. Space missions such as CoRoT
and {\it Kepler} are anticipated to provide many examples of
long-period eclipsing binaries with late-type components, although
they may be faint.

This is an area where long-baseline interferometry has already made
and continues to make important contributions.  While it may be
difficult to find suitable long-period eclipsing binaries with
late-type components in order to measure their radii, single stars of
late spectral type are plentiful, and more easily studied. In recent
years several groups (PTI, VLTI, CHARA) have measured very precise
angular diameters for late-type stars with known parallaxes
\citep[][and others]{Lane:01, Pijpers:03, Segransan:03, DiFolco:04,
Berger:06, Kervella:08, Baines:08, Boyajian:08, Demory:09}. In some
cases the precision obtained for the absolute radii is quite
competitive with that achieved in eclipsing binary systems. And while
there are no dynamical masses to accompany the interferometric radii,
it is usually possible to make use of near-infrared mass-luminosity
relations such as those by \cite{Delfosse:00} to infer sufficiently
precise mass estimates for the purpose of placing the stars on the
mass-radius diagram and comparing with theory.

\cite{Demory:09} have made such a comparison, and report finding that
interferometrically measured stars (which are typically rotating
slowly, and are thus presumably relatively inactive) do in fact agree
with stellar evolution models much better than stars in eclipsing
binaries, as one might expect. Other groups reach a different
conclusion, however, so a final answer must await further
interferometric observations \citep[see][]{Boyajian:10}.

Stars with interferometrically measured radii good to better than 3\%
are indicated with open circles in the bottom panel of
Figure~\ref{fig:mr}.  As was the case with the masses, interferometry
is seen to be very complementary to the eclipsing binaries in
providing accurate radii for the lowest-mass systems.

\subsection{Properties of metal-poor stars}

One of the areas in which stellar evolution models are most poorly
constrained is that of stars with chemical compositions very different
from the Sun, and in particular those that are metal-poor. Detached
binary stars that are suitable for accurate mass and radius
determinations and that have known metal-poor compositions are rare.
Eclipsing binaries in globular clusters are of course obvious targets,
but they tend to be faint and difficult to study. One of the best
examples published recently is that of the variable V69 in 47~Tucanae
\citep{Thompson:10}, with an adopted metallicity for the cluster of
[Fe/H] $= -0.70$ and $\alpha$-element enhancement of [$\alpha$/Fe] $=
+0.4$. Possibly the most metal-poor field eclipsing binary with
accurately known masses and radii is V432~Aur \citep{Siviero:04}, with
a measured [Fe/H] $= -0.60$. Other binary candidates in clusters with
much lower metallicity are known \citep[e.g.,][]{Thompson:01,
Kaluzny:06, Kaluzny:08}, but their properties are not yet determined
sufficiently accurately to be useful for testing models.

Once again long-baseline interferometry has an advantage, as it does
not require the binary to be eclipsing in order to determine the
dynamical masses of its components. A good example is HD~195987
\citep{Torres:02b}, a nearby high proper motion field star with [Fe/H]
$= -0.50$ and [$\alpha$/Fe] $= +0.36$, in which the masses of both
stars were determined to better than 2\% based on interferometric
measurements with the PTI.

\subsection{Properties of evolved stars}

Binary stars with components in rapid phases of evolution (giants or
subgiants) that are sufficiently detached so that they don't interfere
with each other and that are suitable for high-precision mass
determinations are also quite rare. Only a handful of eclipsing
systems of this type are known, including AI~Phe, TZ~For, and the
system OGLE~051019.64$-$685812.3 in the LMC \citep{Andersen:88,
Andersenetal:91, Imbert:87, Pietrzynski:09}. Astrometric-spectroscopic
systems that are amenable to interferometric studies are somewhat more
common, but not many have yielded the precision needed for testing
models. One such example investigated with the PTI is HD~9939
\citep{Boden:06}, in which the primary star appears to be traversing
the Hertzprung gap.

Perhaps the most prominent example of a pair of giants studied
interferometrically is Capella ($\alpha$~Aur, G8\,III + G0\,III),
which is in fact the very first system studied with this technique
\citep{Anderson:20, Merrill:22} using the original Michelson
interferometer on Mount Wilson. Despite its century-long observational
history, the properties of Capella have been surprisingly difficult to
pin down, particularly the masses. The recent study by
\cite{Torres:09} made use of all available interferometric data (from
COAST, Mark~III, IOTA, and other instruments) combined with radial
velocity measurements spanning more than 100 years, and derived
component masses with formal errors smaller than 0.7\%. Capella is
unique among the evolved systems in the amount of information
available for the system, which in addition to the masses includes the
absolute radii (from angular diameter measurements), spectroscopic
effective temperatures, independently determined luminosities (based
on the accurate orbital parallax), projected rotational velocities $v
\sin i$, rotational periods from chromospheric activity indicators,
and importantly, the chemical composition. The latter includes not
only the overall metallicity [m/H], but also the carbon isotope ratio
$^{12}$C/$^{13}$C for the primary, and the lithium abundance and
carbon-to-nitrogen ratios for both stars. The last three quantities
are sensitive diagnostics of evolution, and change drastically for
giants as a result of the deepening of the convective envelope during
the first dredge-up.

The secondary component is crossing the Hertzprung gap, while the
primary is believed to be in the longer-lived phase in which it burns
helium in the core (``clump giant''), although the observational
evidence for this is still somewhat controversial.  With so much
information one would think that the precise evolutionary state of the
primary ought to be very well established. However, the study by
\cite{Torres:09} concluded that current models of stellar evolution
are unable to fit all observational constraints simultaneously for
both stars, at a single age. Very recently a new spectroscopic study
by \cite{Weber:11} has yielded masses with even smaller formal errors
of about 0.3\%, but which differ from the previous values by 4\% and
2\% for the primary and secondary, respectively.  Thus, the last word
is yet to be said on the evolutionary state of the primary and the
ability of current models to match all observational constraints;
Capella is not yielding its secrets so easily.

\subsection{Interferometry and the Pleiades distance}

Soon after the publication of the Hipparcos results on the
trigonometric parallaxes of $\sim$118,000 stars, a controversy ensued
regarding the distance to the Pleiades cluster. Based on measurements
for 55 member stars observed by the satellite in this cluster,
\cite{vanLeeuwen:99} reported an average distance of $118.3 \pm
3.5$~pc, which was in disagreement with results based on the widely
used method of main-sequence isochrone fitting. That technique gave
significantly larger values \citep[e.g., $131.8 \pm
2.5$~pc;][]{Pinsonneault:98}. The difference of $\sim$10\% corresponds
to about 1~mas in the average parallax of the cluster, or 0.23~mag in
the distance modulus. The prospect of systematic errors at this level
in the Hipparcos parallaxes was a rather serious concern and was
difficult to accept for some, particularly since no such problem had
been detected in the parallaxes of other open clusters including the
Hyades. But the possibility that stellar evolution models, which had
worked so well in the past, could be off by as much as 0.23~mag was
equally worrisome, and could have wide-ranging implications for much
of Astrophysics.

A partial solution to the problem came from several different fronts,
one of them involving long-baseline interferometry of a binary system
in the cluster.  \cite{Zwahlen:04} made use of interferometric
observations of the bright 291-day B8\,III binary star Atlas (27~Tau,
HD~23850) collected with the Mark~III interferometer and with NPOI,
and for the first time measured also radial velocities of both
components (only the primary had been measured previously), with which
they derived the spectroscopic orbit. From these measurements they
obtained the orbital parallax, corresponding to a distance of $132 \pm
4$~pc, supporting the isochrone fitting value. Additional support for
this larger distance came from another binary in the cluster, the
double-lined eclipsing system HD~23642, in which a determination of
the absolute properties (particularly $R$ and $T_{\rm eff}$, and
therefore $L$) led to a largely model-independent distance estimate
\citep{Torres:03, Munari:04, Southworth:05, Groenewegen:07}. Further
support came from directly measured trigonometric parallaxes of three
cluster members using the Fine Guidance Sensors on HST
\citep{Soderblom:05}. A re-reduction of the Hipparcos data
\citep{vanLeeuwen:07} removed correlated errors that affected the
original parallaxes, and reduced the discrepancy somewhat (giving a
distance of $122.2 \pm 2.0$~pc), but not completely.  The remaining
difference has not been explained; it may have something to do with
the depth of the cluster and the particular location of the stars
studied, or other unrecognized systematic biases \citep[for a review
on the subject, see][]{Perryman:09}.

\subsection{Towards higher precision in binary masses}

Typical uncertainties in the absolute mass determinations for the best
studied eclipsing binaries are 1--2\%, with a few systems reaching
values as low as a few tenths of a percent. Recent work by
\cite{Konacki:10} has attempted to push these limits even further, for
selected double-lined spectroscopic-interferometric binaries
\citep[see also earlier work focusing on eclipsing binaries
by][]{Lacy:92}. The improvements are based in part on a spectroscopic
technique borrowed from the exoplanet search programs for measuring
very precise radial velocities of stars, applied here to composite
spectra rather than to single-lined spectra \citep{Konacki:05}.  The
method uses an iodine cell in front of the spectrograph slit to track
spectrograph drifts and changes in the point-spread function that
normally lead to systematic errors in the radial velocities
\citep{Marcy:92, Butler:96}. The precision achievable in the
velocities for these double-lined systems is a few tens of
m~s$^{-1}$. Interferometric orbits for the binaries in the work of
\cite{Konacki:10} were obtained using the PTI, with emphasis placed on
those with nearly edge-on orbits. This maximizes the precision in the
masses, given that in astrometric-spectroscopic systems the mass $M$
is inversely proportional to $\sin^3 i$, where $i$ is the inclination
angle of the orbit:
\begin{eqnarray*}
M_1 \sin^3 i & = & P(1-e^2)^{3/2} (K_1+K_2)^2 K_2 / 2\pi G~~ \\ 
M_2 \sin^3 i & = & P(1-e^2)^{3/2} (K_1+K_2)^2 K_1 / 2\pi G~.
\end{eqnarray*}
In the above expressions $P$ is the orbital period, $e$ the
eccentricity, and $K_1$ and $K_2$ are the semi-amplitudes of the
radial velocity curves.  For example, if the inclination angle of the
orbit is 10\deg\ (nearly face-on) and one wishes to obtain a relative
precision of 3\% in the masses, the precision in $i$ must be at least
0\fdg1 in order for the astrometry not to dominate the error budget.
Such a small error in $i$ can be difficult to achieve.  However, if
the binary has an inclination of 87\deg\ (i.e., almost edge-on), one
can get away with an uncertainty in $i$ as large as 10\deg\ (100 times
worse than before), and still be able to measure masses to 3\%
precision.

Two of the binary systems studied by \cite{Konacki:10} reach record
precision in the masses of the components. The formal errors for the
F-star system HD~123999 are 0.20\%, while for HD~210027 they are as
low as 0.066\% (the smallest error obtained for any normal star).
Orbital parallaxes for these two binaries were also obtained to very
high precision, the uncertainties being only 44 and 32 micro arc
seconds, respectively ($\sim$0.15\%). The precision of the masses for
HD~210027 rivals that of the best known determinations in double
neutron star systems, measured by radio pulsar timing.
\cite{Konacki:10} expect that other binary stars may yield similar
precisions if they are selected to have favorable properties,
including masses between 0.5~$M_{\odot}$ and 1.5~$M_{\odot}$ so that
they have sufficiently numerous and sharp spectral lines, periods
between 3 and 23 days, inclination angles between 85\deg\ and 90\deg,
uncertainties in $i$ no larger than 0\fdg3, and errors in the
radial-velocity semi-amplitudes under 31~m~s$^{-1}$.

\section{Final remarks}

Long-baseline interferometry continues to make important contributions
to our knowledge of accurate masses and other fundamental properties
of stars. However, it is a scarce resource: there are not many of
these instruments in the world, and it is generally a very difficult
technique that requires specialized skills. It also has significant
limitations regarding sensitivity (although this is improving, and
there are high hopes for the Magdalena Ridge Observatory
Interferometer currently under construction). Consequently, it is
important to use these facilities wisely. 

Astronomy would benefit the most from applications of interferometry
to objects of special astrophysical interest, rather than those that
are easiest to observe. In the field of accurate fundamental
parameters of stars, there are several areas of the H-R diagram where
much work remains to be done in order to constrain models of stellar
evolution. This includes low-mass stars, high-mass stars, evolved
stars (giants and subgiants), pre-main sequence stars, and stars of
non-solar metallicity. As described above for some of these
categories, constraints on the models are either very scarce, or
problems with theory have already been identified that require
additional observations to help guide theorists toward a better
understanding of the discrepancies. When considering recent
improvements in observational and analysis techniques, prospects are
good for significantly increased precision in fundamental stellar
properties.

\acknowledgements This work was partially supported by NSF grant
AST-1007992. The author is grateful to the meeting organizers for the
invitation to present this paper and for their travel support, as well
as the opportunity for fruitful interaction with other researchers in
this field.

\bibliography{paper}

\begin{thebibliography}{}
\expandafter\ifx\csname natexlab\endcsname\relax\def\natexlab#1{#1}\fi
\expandafter\ifx\csname url\endcsname\relax
  \def\url#1{\texttt{#1}}\fi
\expandafter\ifx\csname urlprefix\endcsname\relax\def\urlprefix{URL }\fi
\providecommand{\eprint}[2][]{\url{#2}}

\bibitem[{{Andersen}(1991)}]{Andersen:91}
{Andersen}, J. 1991, \aapr, 3, 91

\bibitem[{{Andersen} et~al.(1988){Andersen}, {Clausen}, {Nordstrom},
  {Gustafsson}, \& {Vandenberg}}]{Andersen:88}
{Andersen}, J., {Clausen}, J.~V., {Nordstrom}, B., {Gustafsson}, B., \&
  {Vandenberg}, D.~A. 1988, \aap, 196, 128

\bibitem[{{Andersen} et~al.(1991){Andersen}, {Clausen}, {Nordstrom}, {Tomkin},
  \& {Mayor}}]{Andersenetal:91}
{Andersen}, J., {Clausen}, J.~V., {Nordstrom}, B., {Tomkin}, J., \& {Mayor}, M.
  1991, \aap, 246, 99

\bibitem[{{Anderson}(1920)}]{Anderson:20}
{Anderson}, J.~A. 1920, \apj, 51, 263

\bibitem[{{Baines} et~al.(2008){Baines}, {McAlister}, {ten Brummelaar},
  {Turner}, {Sturmann}, {Sturmann}, {Goldfinger}, \& {Ridgway}}]{Baines:08}
{Baines}, E.~K., {McAlister}, H.~A., {ten Brummelaar}, T.~A., {Turner}, N.~H.,
  {Sturmann}, J., {Sturmann}, L., {Goldfinger}, P.~J., \& {Ridgway}, S.~T.
  2008, \apj, 680, 728. \eprint{0803.1411}

\bibitem[{{Berger} et~al.(2006){Berger}, {Gies}, {McAlister}, {ten Brummelaar},
  {Henry}, {Sturmann}, {Sturmann}, {Turner}, {Ridgway}, {Aufdenberg}, \&
  {M{\'e}rand}}]{Berger:06}
{Berger}, D.~H., {Gies}, D.~R., {McAlister}, H.~A., {ten Brummelaar}, T.~A.,
  {Henry}, T.~J., {Sturmann}, J., {Sturmann}, L., {Turner}, N.~H., {Ridgway},
  S.~T., {Aufdenberg}, J.~P., \& {M{\'e}rand}, A. 2006, \apj, 644, 475.
  \eprint{arXiv:astro-ph/0602105}

\bibitem[{{Blackwell} \& {Shallis}(1977)}]{Blackwell:77}
{Blackwell}, D.~E., \& {Shallis}, M.~J. 1977, \mnras, 180, 177

\bibitem[{{Blackwell} et~al.(1979){Blackwell}, {Shallis}, \&
  {Selby}}]{Blackwell:79}
{Blackwell}, D.~E., {Shallis}, M.~J., \& {Selby}, M.~J. 1979, \mnras, 188, 847

\bibitem[{{Boden} et~al.(2006){Boden}, {Torres}, \& {Latham}}]{Boden:06}
{Boden}, A.~F., {Torres}, G., \& {Latham}, D.~W. 2006, \apj, 644, 1193.
  \eprint{arXiv:astro-ph/0601515}

\bibitem[{{Boyajian} et~al.(2008){Boyajian}, {McAlister}, {Baines}, {Gies},
  {Henry}, {Jao}, {O'Brien}, {Raghavan}, {Touhami}, {ten Brummelaar},
  {Farrington}, {Goldfinger}, {Sturmann}, {Sturmann}, {Turner}, \&
  {Ridgway}}]{Boyajian:08}
{Boyajian}, T.~S., {McAlister}, H.~A., {Baines}, E.~K., {Gies}, D.~R., {Henry},
  T., {Jao}, W.-C., {O'Brien}, D., {Raghavan}, D., {Touhami}, Y., {ten
  Brummelaar}, T.~A., {Farrington}, C., {Goldfinger}, P.~J., {Sturmann}, L.,
  {Sturmann}, J., {Turner}, N.~H., \& {Ridgway}, S. 2008, \apj, 683, 424.
  \eprint{0804.2719}

\bibitem[{{Boyajian} et~al.(2010){Boyajian}, {von Braun}, {van Belle}, {ten
  Brummelaar}, {Ciardi}, {Farrington}, {Goldfinger}, {Henry},
  {L{\'o}pez-Morales}, {McAlister}, {Ridgway}, {Schaefer}, {Sturmann},
  {Sturmann}, \& {Turner}}]{Boyajian:10}
{Boyajian}, T.~S., {von Braun}, K., {van Belle}, G., {ten Brummelaar}, T.,
  {Ciardi}, D., {Farrington}, C., {Goldfinger}, P., {Henry}, T.,
  {L{\'o}pez-Morales}, M., {McAlister}, H., {Ridgway}, S.~T., {Schaefer}, G.,
  {Sturmann}, J., {Sturmann}, L., \& {Turner}, N. 2010, in American
  Astronomical Society Meeting Abstracts \#216, vol. 216 of American
  Astronomical Society Meeting Abstracts, \#423.06

\bibitem[{{Butler} et~al.(1996){Butler}, {Marcy}, {Williams}, {McCarthy},
  {Dosanjh}, \& {Vogt}}]{Butler:96}
{Butler}, R.~P., {Marcy}, G.~W., {Williams}, E., {McCarthy}, C., {Dosanjh}, P.,
  \& {Vogt}, S.~S. 1996, \pasp, 108, 500

\bibitem[{{Casagrande} et~al.(2008){Casagrande}, {Flynn}, \&
  {Bessell}}]{Casagrande:08}
{Casagrande}, L., {Flynn}, C., \& {Bessell}, M. 2008, \mnras, 389, 585.
  \eprint{0806.2471}

\bibitem[{{Casagrande} et~al.(2010){Casagrande}, {Ram{\'{\i}}rez},
  {Mel{\'e}ndez}, {Bessell}, \& {Asplund}}]{Casagrande:10}
{Casagrande}, L., {Ram{\'{\i}}rez}, I., {Mel{\'e}ndez}, J., {Bessell}, M., \&
  {Asplund}, M. 2010, \aap, 512, A54+. \eprint{1001.3142}

\bibitem[{{Chabrier} et~al.(2007){Chabrier}, {Gallardo}, \&
  {Baraffe}}]{Chabrier:07}
{Chabrier}, G., {Gallardo}, J., \& {Baraffe}, I. 2007, \aap, 472, L17.
  \eprint{0707.1792}

\bibitem[{{Clausen} et~al.(1999){Clausen}, {Baraffe}, {Claret}, \&
  {Vandenberg}}]{Clausen:99}
{Clausen}, J.~V., {Baraffe}, I., {Claret}, A., \& {Vandenberg}, D.~A. 1999, in
  Stellar Structure: Theory and Test of Connective Energy Transport, edited by
  {A.~Gimenez, E.~F.~Guinan, \& B.~Montesinos}, vol. 173 of Astronomical
  Society of the Pacific Conference Series, 265

\bibitem[{{D'Antona} et~al.(2000){D'Antona}, {Ventura}, \&
  {Mazzitelli}}]{DAntona:00}
{D'Antona}, F., {Ventura}, P., \& {Mazzitelli}, I. 2000, \apjl, 543, L77

\bibitem[{{Delfosse} et~al.(2000){Delfosse}, {Forveille}, {S{\'e}gransan},
  {Beuzit}, {Udry}, {Perrier}, \& {Mayor}}]{Delfosse:00}
{Delfosse}, X., {Forveille}, T., {S{\'e}gransan}, D., {Beuzit}, J.-L., {Udry},
  S., {Perrier}, C., \& {Mayor}, M. 2000, \aap, 364, 217.
  \eprint{arXiv:astro-ph/0010586}

\bibitem[{{Demory} et~al.(2009){Demory}, {S{\'e}gransan}, {Forveille},
  {Queloz}, {Beuzit}, {Delfosse}, {di Folco}, {Kervella}, {Le Bouquin},
  {Perrier}, {Benisty}, {Duvert}, {Hofmann}, {Lopez}, \& {Petrov}}]{Demory:09}
{Demory}, B.-O., {S{\'e}gransan}, D., {Forveille}, T., {Queloz}, D., {Beuzit},
  J.-L., {Delfosse}, X., {di Folco}, E., {Kervella}, P., {Le Bouquin}, J.-B.,
  {Perrier}, C., {Benisty}, M., {Duvert}, G., {Hofmann}, K.-H., {Lopez}, B., \&
  {Petrov}, R. 2009, \aap, 505, 205. \eprint{0906.0602}

\bibitem[{{Di Folco} et~al.(2004){Di Folco}, {Th{\'e}venin}, {Kervella},
  {Domiciano de Souza}, {Coud{\'e} du Foresto}, {S{\'e}gransan}, \&
  {Morel}}]{DiFolco:04}
{Di Folco}, E., {Th{\'e}venin}, F., {Kervella}, P., {Domiciano de Souza}, A.,
  {Coud{\'e} du Foresto}, V., {S{\'e}gransan}, D., \& {Morel}, P. 2004, \aap,
  426, 601

\bibitem[{{Groenewegen} et~al.(2007){Groenewegen}, {Decin}, {Salaris}, \& {De
  Cat}}]{Groenewegen:07}
{Groenewegen}, M.~A.~T., {Decin}, L., {Salaris}, M., \& {De Cat}, P. 2007,
  \aap, 463, 579

\bibitem[{{Hershey} \& {Taff}(1998)}]{Hershey:98}
{Hershey}, J.~L., \& {Taff}, L.~G. 1998, \aj, 116, 1440

\bibitem[{{Hillenbrand} \& {White}(2004)}]{Hillenbrand:04}
{Hillenbrand}, L.~A., \& {White}, R.~J. 2004, \apj, 604, 741.
  \eprint{arXiv:astro-ph/0312189}

\bibitem[{{Holmberg} et~al.(2007){Holmberg}, {Nordstr{\"o}m}, \&
  {Andersen}}]{Holmberg:07}
{Holmberg}, J., {Nordstr{\"o}m}, B., \& {Andersen}, J. 2007, \aap, 475, 519.
  \eprint{0707.1891}

\bibitem[{{Hoxie}(1973)}]{Hoxie:73}
{Hoxie}, D.~T. 1973, \aap, 26, 437

\bibitem[{{Imbert}(1987)}]{Imbert:87}
{Imbert}, M. 1987, \aaps, 71, 69

\bibitem[{{Kaluzny} et~al.(2006){Kaluzny}, {Pych}, {Rucinski}, \&
  {Thompson}}]{Kaluzny:06}
{Kaluzny}, J., {Pych}, W., {Rucinski}, S.~M., \& {Thompson}, I.~B. 2006, Acta
  Astron., 56, 237. \eprint{arXiv:astro-ph/0609380}

\bibitem[{{Kaluzny} et~al.(2008){Kaluzny}, {Thompson}, {Rucinski}, \&
  {Krzeminski}}]{Kaluzny:08}
{Kaluzny}, J., {Thompson}, I.~B., {Rucinski}, S.~M., \& {Krzeminski}, W. 2008,
  \aj, 136, 400. \eprint{0804.4351}

\bibitem[{{Kervella} et~al.(2008){Kervella}, {M{\'e}rand}, {Pichon},
  {Th{\'e}venin}, {Heiter}, {Bigot}, {ten Brummelaar}, {McAlister}, {Ridgway},
  {Turner}, {Sturmann}, {Sturmann}, {Goldfinger}, \&
  {Farrington}}]{Kervella:08}
{Kervella}, P., {M{\'e}rand}, A., {Pichon}, B., {Th{\'e}venin}, F., {Heiter},
  U., {Bigot}, L., {ten Brummelaar}, T.~A., {McAlister}, H.~A., {Ridgway},
  S.~T., {Turner}, N., {Sturmann}, J., {Sturmann}, L., {Goldfinger}, P.~J., \&
  {Farrington}, C. 2008, \aap, 488, 667. \eprint{0806.4049}

\bibitem[{{Konacki}(2005)}]{Konacki:05}
{Konacki}, M. 2005, \apj, 626, 431. \eprint{arXiv:astro-ph/0410389}

\bibitem[{{Konacki} et~al.(2010){Konacki}, {Muterspaugh}, {Kulkarni}, \&
  {He{\l}miniak}}]{Konacki:10}
{Konacki}, M., {Muterspaugh}, M.~W., {Kulkarni}, S.~R., \& {He{\l}miniak},
  K.~G. 2010, \apj, 719, 1293. \eprint{0910.4482}

\bibitem[{{Lacy}(1977)}]{Lacy:77}
{Lacy}, C.~H. 1977, \apjs, 34, 479

\bibitem[{{Lacy}(1992)}]{Lacy:92}
--- 1992, in IAU Colloq. 135: Complementary Approaches to Double and Multiple
  Star Research, edited by {H.~A.~McAlister \& W.~I.~Hartkopf}, vol.~32 of
  Astronomical Society of the Pacific Conference Series, 152

\bibitem[{{Lane} et~al.(2001){Lane}, {Boden}, \& {Kulkarni}}]{Lane:01}
{Lane}, B.~F., {Boden}, A.~F., \& {Kulkarni}, S.~R. 2001, \apjl, 551, L81

\bibitem[{{Lastennet} \& {Valls-Gabaud}(2002)}]{Lastennet:02}
{Lastennet}, E., \& {Valls-Gabaud}, D. 2002, \aap, 396, 551.
  \eprint{arXiv:astro-ph/0211501}

\bibitem[{{L{\'o}pez-Morales}(2007)}]{Lopez-Morales:07}
{L{\'o}pez-Morales}, M. 2007, \apj, 660, 732. \eprint{arXiv:astro-ph/0701702}

\bibitem[{{L{\'o}pez-Morales} \& {Ribas}(2005)}]{Lopez-Morales:05}
{L{\'o}pez-Morales}, M., \& {Ribas}, I. 2005, \apj, 631, 1120.
  \eprint{arXiv:astro-ph/0505001}

\bibitem[{{Marcy} \& {Butler}(1992)}]{Marcy:92}
{Marcy}, G.~W., \& {Butler}, R.~P. 1992, \pasp, 104, 270

\bibitem[{{Merrill}(1922)}]{Merrill:22}
{Merrill}, P.~W. 1922, \apj, 56, 40

\bibitem[{{Morales} et~al.(2010){Morales}, {Gallardo}, {Ribas}, {Jordi},
  {Baraffe}, \& {Chabrier}}]{Morales:10}
{Morales}, J.~C., {Gallardo}, J., {Ribas}, I., {Jordi}, C., {Baraffe}, I., \&
  {Chabrier}, G. 2010, \apj, 718, 502. \eprint{1005.5720}

\bibitem[{{Morales} et~al.(2008){Morales}, {Ribas}, \& {Jordi}}]{Morales:08}
{Morales}, J.~C., {Ribas}, I., \& {Jordi}, C. 2008, \aap, 478, 507.
  \eprint{0711.3523}

\bibitem[{{Morales} et~al.(2009){Morales}, {Ribas}, {Jordi}, {Torres},
  {Gallardo}, {Guinan}, {Charbonneau}, {Wolf}, {Latham}, {Anglada-Escud{\'e}},
  {Bradstreet}, {Everett}, {O'Donovan}, {Mandushev}, \& {Mathieu}}]{Morales:09}
{Morales}, J.~C., {Ribas}, I., {Jordi}, C., {Torres}, G., {Gallardo}, J.,
  {Guinan}, E.~F., {Charbonneau}, D., {Wolf}, M., {Latham}, D.~W.,
  {Anglada-Escud{\'e}}, G., {Bradstreet}, D.~H., {Everett}, M.~E., {O'Donovan},
  F.~T., {Mandushev}, G., \& {Mathieu}, R.~D. 2009, \apj, 691, 1400.
  \eprint{0810.1541}

\bibitem[{{Mullan} \& {MacDonald}(2001)}]{Mullan:01}
{Mullan}, D.~J., \& {MacDonald}, J. 2001, \apj, 559, 353

\bibitem[{{Munari} et~al.(2004){Munari}, {Dallaporta}, {Siviero}, {Soubiran},
  {Fiorucci}, \& {Girard}}]{Munari:04}
{Munari}, U., {Dallaporta}, S., {Siviero}, A., {Soubiran}, C., {Fiorucci}, M.,
  \& {Girard}, P. 2004, \aap, 418, L31. \eprint{arXiv:astro-ph/0403444}

\bibitem[{{Perryman}(2009)}]{Perryman:09}
{Perryman}, M. 2009, {Astronomical Applications of Astrometry: Ten Years of
  Exploitation of the Hipparcos Satellite Data} (Cambridge University Press)

\bibitem[{{Perryman} et~al.(1997){Perryman}, {Lindegren}, {Kovalevsky}, {Hoeg},
  {Bastian}, {Bernacca}, {Cr{\'e}z{\'e}}, {Donati}, {Grenon}, {van Leeuwen},
  {van der Marel}, {Mignard}, {Murray}, {Le Poole}, {Schrijver}, {Turon},
  {Arenou}, {Froeschl{\'e}}, \& {Petersen}}]{Perryman:97}
{Perryman}, M.~A.~C., {Lindegren}, L., {Kovalevsky}, J., {Hoeg}, E., {Bastian},
  U., {Bernacca}, P.~L., {Cr{\'e}z{\'e}}, M., {Donati}, F., {Grenon}, M., {van
  Leeuwen}, F., {van der Marel}, H., {Mignard}, F., {Murray}, C.~A., {Le
  Poole}, R.~S., {Schrijver}, H., {Turon}, C., {Arenou}, F., {Froeschl{\'e}},
  M., \& {Petersen}, C.~S. 1997, \aap, 323, L49

\bibitem[{{Pietrzy{\'n}ski} et~al.(2009){Pietrzy{\'n}ski}, {Thompson},
  {Graczyk}, {Gieren}, {Udalski}, {Szewczyk}, {Minniti}, {Ko{\l}aczkowski},
  {Bresolin}, \& {Kudritzki}}]{Pietrzynski:09}
{Pietrzy{\'n}ski}, G., {Thompson}, I.~B., {Graczyk}, D., {Gieren}, W.,
  {Udalski}, A., {Szewczyk}, O., {Minniti}, D., {Ko{\l}aczkowski}, Z.,
  {Bresolin}, F., \& {Kudritzki}, R.-P. 2009, \apj, 697, 862.
  \eprint{0903.0855}

\bibitem[{{Pijpers} et~al.(2003){Pijpers}, {Teixeira}, {Garcia}, {Cunha},
  {Monteiro}, \& {Christensen-Dalsgaard}}]{Pijpers:03}
{Pijpers}, F.~P., {Teixeira}, T.~C., {Garcia}, P.~J., {Cunha}, M.~S.,
  {Monteiro}, M.~J.~P.~F.~G., \& {Christensen-Dalsgaard}, J. 2003, \aap, 406,
  L15

\bibitem[{{Pinsonneault} et~al.(1998){Pinsonneault}, {Stauffer}, {Soderblom},
  {King}, \& {Hanson}}]{Pinsonneault:98}
{Pinsonneault}, M.~H., {Stauffer}, J., {Soderblom}, D.~R., {King}, J.~R., \&
  {Hanson}, R.~B. 1998, \apj, 504, 170. \eprint{arXiv:astro-ph/9803233}

\bibitem[{{Pols} et~al.(1997){Pols}, {Tout}, {Schroder}, {Eggleton}, \&
  {Manners}}]{Pols:97}
{Pols}, O.~R., {Tout}, C.~A., {Schroder}, K.-P., {Eggleton}, P.~P., \&
  {Manners}, J. 1997, \mnras, 289, 869

\bibitem[{{Popper}(1967)}]{Popper:67}
{Popper}, D.~M. 1967, \araa, 5, 85

\bibitem[{{Popper}(1980)}]{Popper:80}
--- 1980, \araa, 18, 115

\bibitem[{{Popper}(1997)}]{Popper:97}
--- 1997, \aj, 114, 1195

\bibitem[{{Ram{\'{\i}}rez} \& {Mel{\'e}ndez}(2005)}]{Ramirez:05}
{Ram{\'{\i}}rez}, I., \& {Mel{\'e}ndez}, J. 2005, \apj, 626, 446.
  \eprint{arXiv:astro-ph/0503108}

\bibitem[{{Ribas}(2003)}]{Ribas:03}
{Ribas}, I. 2003, \aap, 398, 239. \eprint{arXiv:astro-ph/0211086}

\bibitem[{{Schlaufman} \& {Laughlin}(2010)}]{Schlaufman:10}
{Schlaufman}, K.~C., \& {Laughlin}, G. 2010, \aap, 519, A105+.
  \eprint{1006.2850}

\bibitem[{{S{\'e}gransan} et~al.(2003){S{\'e}gransan}, {Kervella}, {Forveille},
  \& {Queloz}}]{Segransan:03}
{S{\'e}gransan}, D., {Kervella}, P., {Forveille}, T., \& {Queloz}, D. 2003,
  \aap, 397, L5. \eprint{arXiv:astro-ph/0211647}

\bibitem[{{Siviero} et~al.(2004){Siviero}, {Munari}, {Sordo}, {Dallaporta},
  {Marrese}, {Zwitter}, \& {Milone}}]{Siviero:04}
{Siviero}, A., {Munari}, U., {Sordo}, R., {Dallaporta}, S., {Marrese}, P.~M.,
  {Zwitter}, T., \& {Milone}, E.~F. 2004, \aap, 417, 1083.
  \eprint{arXiv:astro-ph/0309691}

\bibitem[{{Soderblom} et~al.(2005){Soderblom}, {Nelan}, {Benedict}, {McArthur},
  {Ramirez}, {Spiesman}, \& {Jones}}]{Soderblom:05}
{Soderblom}, D.~R., {Nelan}, E., {Benedict}, G.~F., {McArthur}, B., {Ramirez},
  I., {Spiesman}, W., \& {Jones}, B.~F. 2005, \aj, 129, 1616.
  \eprint{arXiv:astro-ph/0412093}

\bibitem[{{Southworth} et~al.(2005){Southworth}, {Maxted}, \&
  {Smalley}}]{Southworth:05}
{Southworth}, J., {Maxted}, P.~F.~L., \& {Smalley}, B. 2005, \aap, 429, 645.
  \eprint{arXiv:astro-ph/0409507}

\bibitem[{{Thompson} et~al.(2001){Thompson}, {Kaluzny}, {Pych}, {Burley},
  {Krzeminski}, {Paczy{\'n}ski}, {Persson}, \& {Preston}}]{Thompson:01}
{Thompson}, I.~B., {Kaluzny}, J., {Pych}, W., {Burley}, G., {Krzeminski}, W.,
  {Paczy{\'n}ski}, B., {Persson}, S.~E., \& {Preston}, G.~W. 2001, \aj, 121,
  3089. \eprint{arXiv:astro-ph/0012493}

\bibitem[{{Thompson} et~al.(2010){Thompson}, {Kaluzny}, {Rucinski},
  {Krzeminski}, {Pych}, {Dotter}, \& {Burley}}]{Thompson:10}
{Thompson}, I.~B., {Kaluzny}, J., {Rucinski}, S.~M., {Krzeminski}, W., {Pych},
  W., {Dotter}, A., \& {Burley}, G.~S. 2010, \aj, 139, 329. \eprint{0910.4262}

\bibitem[{{Torres}(2003)}]{Torres:03}
{Torres}, G. 2003, Information Bulletin on Variable Stars, 5402, 1

\bibitem[{{Torres}(2004)}]{Torres:04}
--- 2004, in Spectroscopically and Spatially Resolving the Components of the
  Close Binary Stars, edited by {R.~W.~Hilditch, H.~Hensberge, \&
  K.~Pavlovski}, vol. 318 of Astronomical Society of the Pacific Conference
  Series, 123. \eprint{arXiv:astro-ph/0312147}

\bibitem[{{Torres}(2010)}]{Torres:10a}
--- 2010, \aj, 140, 1158. \eprint{1008.3913}

\bibitem[{{Torres} et~al.(2010){Torres}, {Andersen}, \&
  {Gim{\'e}nez}}]{Torres:10b}
{Torres}, G., {Andersen}, J., \& {Gim{\'e}nez}, A. 2010, \aapr, 18, 67.
  \eprint{0908.2624}

\bibitem[{{Torres} et~al.(2002){Torres}, {Boden}, {Latham}, {Pan}, \&
  {Stefanik}}]{Torres:02b}
{Torres}, G., {Boden}, A.~F., {Latham}, D.~W., {Pan}, M., \& {Stefanik}, R.~P.
  2002, \aj, 124, 1716

\bibitem[{{Torres} et~al.(2009){Torres}, {Claret}, \& {Young}}]{Torres:09}
{Torres}, G., {Claret}, A., \& {Young}, P.~A. 2009, \apj, 700, 1349.
  \eprint{0906.0977}

\bibitem[{{Torres} \& {Ribas}(2002)}]{Torres:02a}
{Torres}, G., \& {Ribas}, I. 2002, \apj, 567, 1140.
  \eprint{arXiv:astro-ph/0111167}

\bibitem[{{Twarog} et~al.(2007){Twarog}, {Vargas}, \&
  {Anthony-Twarog}}]{Twarog:07}
{Twarog}, B.~A., {Vargas}, L.~C., \& {Anthony-Twarog}, B.~J. 2007, \aj, 134,
  1777. \eprint{0707.4446}

\bibitem[{{van Leeuwen}(1999)}]{vanLeeuwen:99}
{van Leeuwen}, F. 1999, \aap, 341, L71

\bibitem[{{van Leeuwen}(2007)}]{vanLeeuwen:07}
--- 2007, \aap, 474, 653. \eprint{0708.1752}

\bibitem[{{Weber} \& {Strassmeier}(2011)}]{Weber:11}
{Weber}, M., \& {Strassmeier}, K.~G. 2011, \aap, 531, A89+. \eprint{1104.0342}

\bibitem[{{Windmiller} et~al.(2010){Windmiller}, {Orosz}, \&
  {Etzel}}]{Windmiller:10}
{Windmiller}, G., {Orosz}, J.~A., \& {Etzel}, P.~B. 2010, \apj, 712, 1003.
  \eprint{1002.2003}

\bibitem[{{Zwahlen} et~al.(2004){Zwahlen}, {North}, {Debernardi}, {Eyer},
  {Galland}, {Groenewegen}, \& {Hummel}}]{Zwahlen:04}
{Zwahlen}, N., {North}, P., {Debernardi}, Y., {Eyer}, L., {Galland}, F.,
  {Groenewegen}, M.~A.~T., \& {Hummel}, C.~A. 2004, \aap, 425, L45.
  \eprint{arXiv:astro-ph/0408430}

\end{thebibliography}

\end{document}